\documentclass{DISproc}

\usepackage{times,fancyhdr,bbold}
\usepackage{cite}
\usepackage{graphicx}
\usepackage{subfig}

\begin{document}

\newcommand{\dif}{{\rm{d}}}
\newcommand{\e}{\rm{e}}
\newcommand{\B}{\mathcal{B}}
\newcommand{\R}{\mathcal{R}}
\newcommand{\K}{\mathcal{K}}
\newcommand{\A}{\mathcal{A}}
\newcommand{\be}{\begin{equation}}
\newcommand{\ee}{\end{equation}}
\newcommand{\bi}{\begin{itemize}}
\newcommand{\ei}{\end{itemize}}
\newcommand{\med}{\medskip \\}
\newcommand{\bq}{\mathbf{q}}
\newcommand{\bk}{\mathbf{k}}
\newcommand{\bl}{\mathbf{l}}
\newcommand{\non}{\nonumber\\}
\newcommand{\asbar}{\bar{\alpha}_s}

%------------------------------------
\title{Using the BFKL resummation to fit DIS data: collinear and running coupling effects}

%for single authors the superscripts are optional
\author{{\slshape M. Hentschinski, A. Sabio Vera, C. Salas}\\[1ex]
Instituto de F{\' \i}sica Te{\' o}rica UAM/CSIC, U. Aut{\' o}noma de Madrid, E-28049 Madrid, Spain}

% please enter the contribution ID for the DOI
\contribID{97}

\doi  % if there is an online version we will register DOIs

\maketitle

\begin{abstract}
The proton structure function $F_2$ is studied in the low $x$ regime using BFKL evolution. The next to leading logarithmic (NLL) analysis requires the inclusion of running coupling effects which lead to off-diagonal terms in the evolution kernel. An all-orders resummation is used to improve the collinear behavior of the NLL BFKL result. We emphasize the theoretical uncertainties that appear throughout the analysis and give a comparison to the combined HERA data.
\end{abstract}

\section{Introduction}

In 2010 HERA made public the combined results~\cite{Aaron:2009aa} obtained by H1~\cite{Ahmed:1995fd} and ZEUS~\cite{Derrick:1995ef} Collaborations for the proton structure function $F_2 (x, Q^2)$ at low values of the Bjorken $x$ variable and a quite broad range of values of the photon virtuality $Q^2$. This observable became specially convenient to test the region of applicability of the theory based on the high energy or Regge limit, which corresponds to the center of mass energy of the system $\sqrt{s}$ being asymptotically larger than any other scale involved. In Deep Inelastic Scattering (DIS) a hard scale is provided by the high virtuality of the photon. Since the $x$ variable is given within very good approximation by the ratio between the photon virtuality and the center of mass energy squared, we can refer to the Regge limit as the limit of low Bjorken $x$. In this regime large logarithms of energy appear, dominating the scattering amplitude, and they need to be resummed to all orders. Such a resummation is achieved by the so-called BFKL evolution equation~\cite{Fadin:1975cb, Lipatov:1976zz, Kuraev:1976ge, Kuraev:1977fs, Balitsky:1978ic}.

The aim of the present study is to analyze the theoretical uncertainties encountered in the determination of $F_2$ at NLL accuracy using high energy factorization~\cite{Catani:1990eg}. Care has to be taken when introducing running coupling effects and it is also needed to resum to all orders the leading collinear singularities which are numerically large in this kinematical region~\cite{Salam:1998tj,Ciafaloni:1999yw,Ciafaloni:2003rd, Altarelli:2005ni,Vera:2005jt}. Concerning the running of the coupling, we compare the results obtained using a running with a Landau pole with a model which freezes in the infrared and is compatible with power corrections to jet observables~\cite{Webber:1998um}. A numerical analysis of the gluon Green function in the kinematic region of interest is provided. We conclude with a preliminary comparison of our theoretical calculation for $F_2$ with HERA data.

\newpage

\section{Analyzing the proton structure function}

\begin{wrapfigure}{l}{0.25\textwidth}
\vspace{-.7cm}
  \centering
  \includegraphics[width=.25\textwidth]{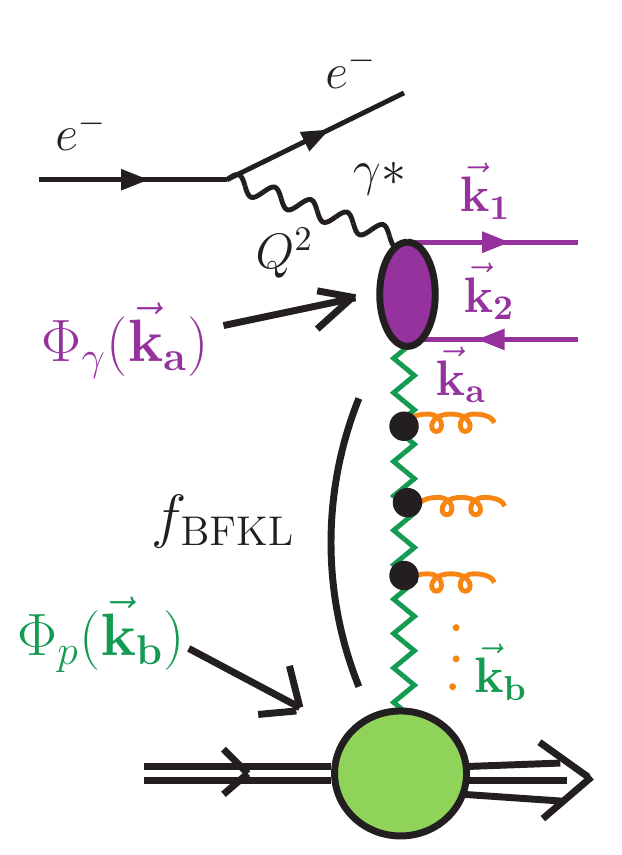}
    \vspace{-.5cm}
  \caption{}
  \label{fig:factorization}
  \vspace{-.4cm}
\end{wrapfigure}

\begin{wrapfigure}{r}{0.4\textwidth}
\vspace{-.5cm}
  \centering
  \includegraphics[width=.4\textwidth]{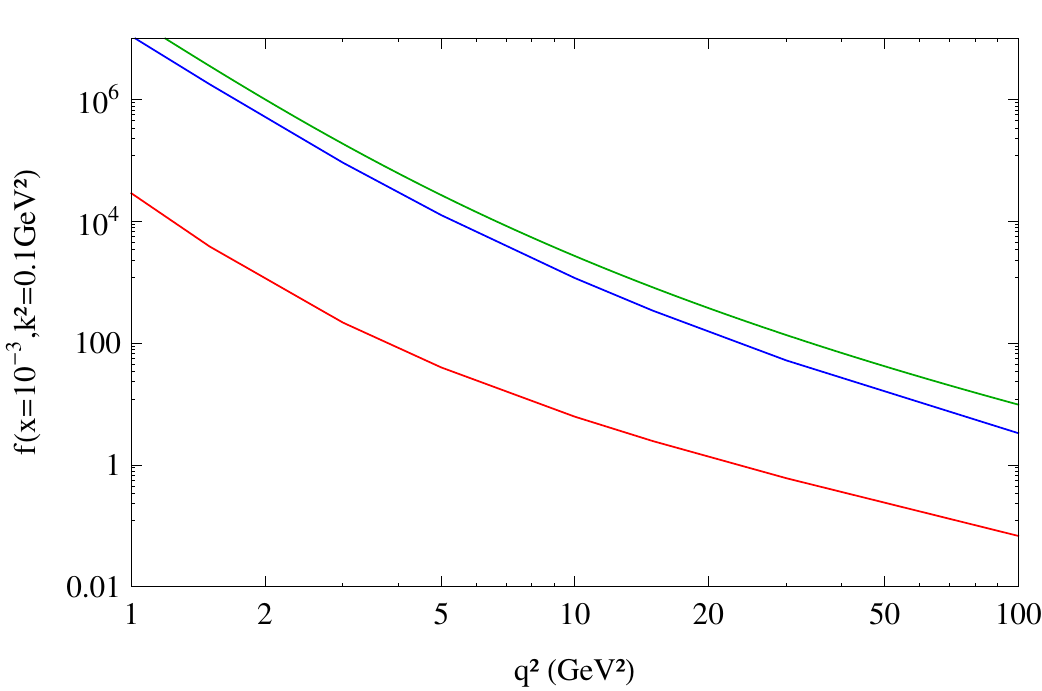}
   % \vspace{-.3cm}
  \caption{Action of the differential operator. Upper curve: LL; mid curve: asymmetric differential operator acting on the photon; and lower curve:  symmetric choice. }
  \label{fig:asysymphot}
 \vspace{-.4cm}
\end{wrapfigure}

\begin{figure}
\vspace{1cm}
\centering
\subfloat[Dependence on $\bk^2$ for $s_0= Q^2$ (solid lines) and $s_0=kq$ (dashed lines) at LL (upper set, green) and NLL collinear improved (lower set, red) accuracy.]{\includegraphics[width=6.5cm,angle=0]{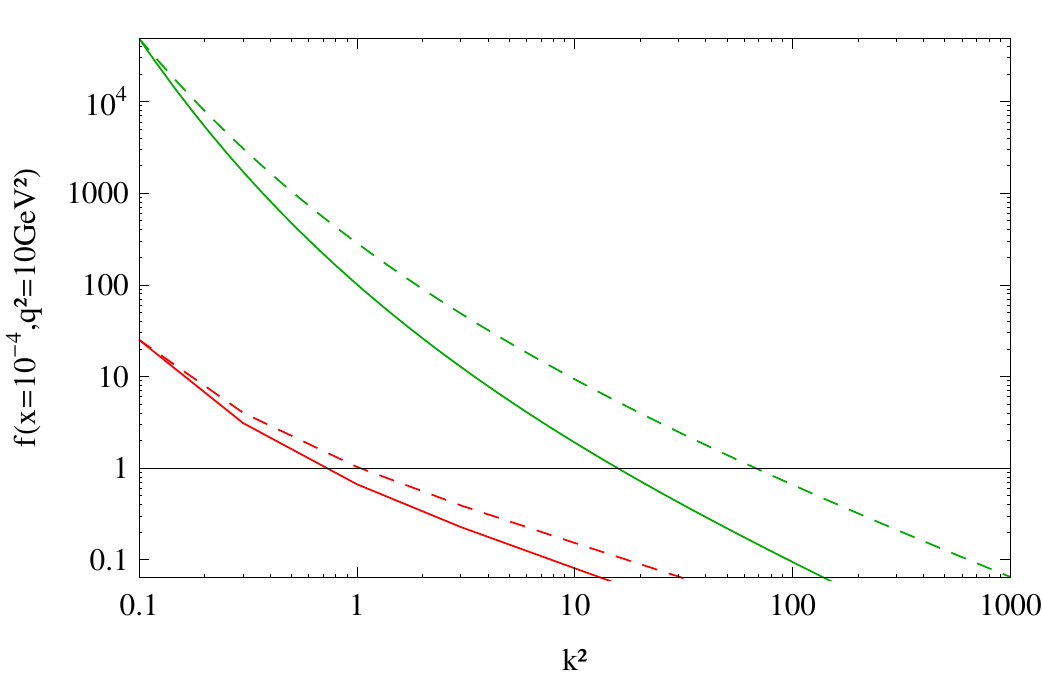} \label{fig:fig21}}\hspace{.3cm}
\subfloat[Dependence on the model for the running: the solid lines correspond to the IR finite running and the dotted ones to the perturbative one at LL (upper set, green) and NLL collinear improved (lower set, red) accuracy.]{\includegraphics[width=6.5cm,angle=0]{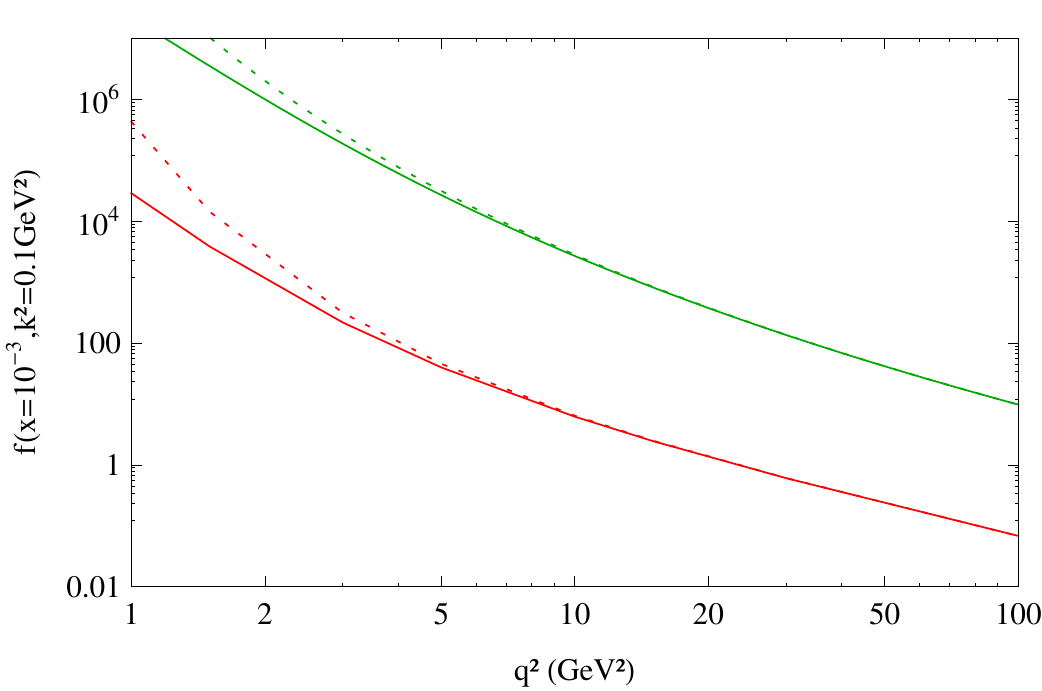} \label{fig:qsym}}
      \caption{Numerical analysis of the gluon Green function.}
   \label{fig:ggf}
\end{figure}

High energy factorization allows to write the proton structure function as a convolution in transverse momentum space of a nonperturbative object describing the proton (proton impact factor $\Phi_p$) with the photon (photon 
impact factor $\Phi_\gamma$), calculated using perturbation theory, together with a gluon Green function $f$, 
linking both process-dependent components and accounting for the BFKL evolution:
\begin{equation*}
F_2(x,Q^2)=\frac{F_c}{(2\pi)^4}\int\frac{\dif^2\bk_a}{\bk^2_a}\int\frac{\dif^2\bk^2_b}{\bk_b}\Phi_{\gamma}(\bk_a)\, f(x,\bk_a,\bk_b)\,\Phi_{p}(\bk_b) \; .
\end{equation*}
Fig.~\ref{fig:factorization} shows the different parts involved. Although an analytic expression for the photon impact factor at next to leading order accuracy~\cite{Fadin:2002tu,Bartels:2002uz,Balitsky:2010ze} has been recently calculated~\cite{chirilli} we use for simplicity in our analysis the leading order result as presented in~\cite{Forshaw:1997dc}. The proton impact factor needs to be modeled. Our results are based on a simple choice which introduces just a few parameters to be determined by the experiment and it presents a Poissonian-like distribution in transverse momentum space with its maximum around the confinement scale. Finally, the gluon Green function is governed by the BFKL equation. Its LL solution is smooth and convergent but not sufficient to explain the DIS data. The first attempt to have a good description of $F_2$ would consist on studying the next order of accuracy. However it is known that the NLL kernel is unstable in collinear regions. We have found that by introducing an all-orders collinear resummation consistent with the NLL solution following the procedure given in~\cite{Vera:2005jt} we not only eliminate the collinear instabilities but also get a good preliminary description of the data. Figs.~\ref{fig:asysymphot} and~\ref{fig:ggf} compare the LL gluon Green function to the collinearly improved one. Since this formalism does not modify the NLL results but only gives higher order corrections there is certain freedom in the way of performing the resummation. In the present analysis we use an expression for the NLL eigenvalues which includes the action on the impact factors of the differential operators breaking the scale invariance of the kernel. 

At NLL accuracy one needs to account for the running of the coupling. This gives an analytical expression for the kernel which contains a differential operator in the Mellin variable $\gamma$~\cite{Kotikov:2000pm}. There is in principle no theoretical restriction (other than having an hermitian hamiltonian) on whether to act with this operator in a symmetric way~\cite{Schwennsen:2007hs}, {\it i.e.} on both proton and photon impact factors or in an asymmetric way (only on one of them). Nevertheless, it turns out that each option produce very different results, as shown in fig.~\ref{fig:asysymphot}. The reason for this is that each of them naturally leads to a different scale for the running coupling due to the scale of the logarithm accompanying the term in $\beta_0$ of the kernel, responsible for the running. This is, however, an assumption about higher order terms again. We could have decided to leave the logarithms without absorbing them into the expression for the running coupling. We have also compared the results obtained for two different models of the running, a perturbative one, with a Landau pole, and the one presented in~\cite{Webber:1998um} and described earlier in the introduction. However, as it can be seen in fig.~\ref{fig:qsym} we are studying a region in which these are minor effects, since we ask the transverse scale to be perturbative. 

The last important point of discussion in this analysis is what to do with the choice of energy scale $s_0$ appearing in the gluon Green function:
\begin{equation*}
f(s,\bk,\bq)=\frac{1}{2\pi^2}\sum_n \int_{-\infty}^\infty\!\!\! \dif \nu \hspace{-.2cm}\int\limits_{\delta-i\infty}\limits^{\delta+i\infty}\frac{\dif \omega}{2\pi i}\, \frac{\e^{i n (\theta_q-\theta_k)}}{\omega-\mathcal{K}(\asbar, 1/2+i\nu)}\,\frac{1}{\bq^2}\left(\frac{\bq^2}{\bk^2}\right)^{1/2+i\nu} \left(\frac{s}{s_{0}}\right)^\omega \; .
\end{equation*}
It is known that any dependence of the cross section on this scale must cancel at NLL accuracy. However, if we want to express $f$ as a function of $x$ a shift in $\omega$ is produced leading to a remaining dependence on it that appear as higher order corrections. A natural choice in this case would be the DIS scale, $s_0 = Q^2$, so that $(s/s_0)^\omega = x^{-\omega}$. The symmetric choice as a product of the internal scales $s_0 = k q$ was used to calculate the NLL BFKL solution~\cite{Kotikov:2000pm}. As it can be seen in fig.~\ref{fig:fig21} there is a difference in the results obtained with each version.

\begin{figure}
\vspace{-1.cm}
  \centering
  \includegraphics[width=.4\textwidth]{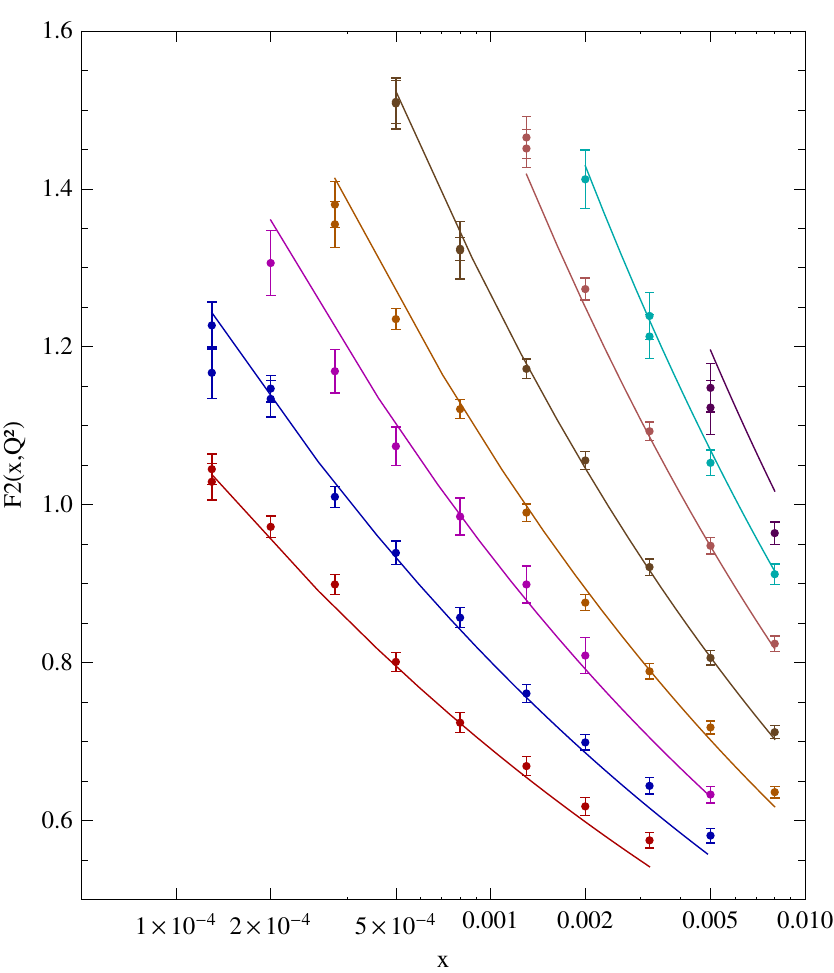}
  %  \vspace{-.3cm}
  \caption{Preliminary fit to $F_2$ with $\delta=1.246$, $Q_0^2=0.368 \text{GeV}^2$ and $A_p=0.07346$. }
  \label{fig:fit}
  \vspace{-.3cm}
\end{figure}

Figure~\ref{fig:fit} shows one of the possible preliminary fits that can be done of $F_2$ within this theoretical framework. In particular, a symmetric version of the differential operator has been used together with the symmetric energy scale $s_0 = k q$, the perturbative running coupling and proton impact factor given by
\begin{equation*}
\Phi_p(\bk^2) = A \left(\bk^2/Q_0^2 \right)^\delta \e^{\bk^2/Q_0^2} \; .
\end{equation*}
The expression for the photon impact factor used can be found in~\cite{Bialas:2001ks}.

A detailed analysis of the work here presented can be found in~\cite{f2}.
 
\section*{Acknowledgements}

The European Comission (LHCPhenoNet PITN-GA-2010-264564) is acknowledged for funding the expenses of the conference.

\newpage

% ****************************************************************************
% BIBLIOGRAPHY AREA
% ****************************************************************************

{\raggedright
\begin{footnotesize}

% ----------------------------------------------------------------------------

% IF YOU USE BIBTEX,
% - DELETE THE TEXT BETWEEN THE TWO ABOVE DASHED LINES
% - UNCOMMENT THE NEXT TWO LINES AND REPLACE 'smith_joe.bib' WITH YOUR
%   FILE(S)

 \bibliographystyle{DISproc}
 \bibliography{citations.bib}
\end{footnotesize}
}

% ****************************************************************************
% END OF BIBLIOGRAPHY AREA
% ****************************************************************************

\end{document}